# TCPTuner: Congestion Control Your Way

Kevin Miller
Stanford University
kmiller4@cs.stanford.edu

Luke Hsiao
Stanford University
lwhsiao@stanford.edu

## ABSTRACT
*TCPTuner* is a TCP (transmission control protocol) congestion control kernel module and GUI (graphical user interface) for Linux that allows real-time modification of the congestion control parameters of TCP CUBIC, the current default algorithm in Linux. Specifically, the tool provides access to *alpha*, the rate at which a sender's congestion window grows; *beta*, the multiplicative factor to decrease the congestion window on a loss event; as well as CUBIC's *fast convergence* and *tcp friendliness* parameters. Additionally, the interface provides access to *ip route* parameters for the minimum retransmission time and initial congestion window size. In this paper, we describe the implementation of *TCPTuner* and show experimental data of the effects of adjusting congestion control parameters.

The source code and instructions to use *TCPTuner* are available at https://github.com/Gasparila/TCPTuner.

## 1. INTRODUCTION
In 1986, the internet faced its first major congestion-based crisis, which resulted in a drastic decrease of productive throughput in the network. Since then, congestion control algorithms have become a fundamental principle of networking and the internet. A study by Lee et. al. suggests that the majority of internet traffic still uses TCP [1], which has a variety of congestion control algorithms that users can employ. For example, Linux includes congestion control algorithms like CUBIC, Vegas, Veno, H-TCP, Westwood, Tahoe, and Reno.

Many of the early congestion control algorithms used an AIMD (additive increase, multiplicative decrease) approach to fully utilize a bottleneck link while converging to a fair allocation of bandwidth [2]. Since then, many other congestion control algorithms have been presented to address different performance challenges. TCP CUBIC, for example, optimizes for high bandwidth, high latency networks [3] and is currently the default congestion control algorithm in Linux and OSX.

Congestion control algorithms are generally thought of as a black box function that users never see nor modify. We believe, however, that users should have control over how their computer operates and be able to easily change the parameters of their TCP congestion control algorithms. To that end, we introduce TCPTuner, a simple software package containing a TCP congestion control kernel module as well as a GUI to directly modify the core parameters.

## 2. RELATED WORK
Tools like *sysctl* and *ip route* exist that allow experienced Linux users to edit kernel parameters. System calls like *setsockopt* also allow programmers to modify a subset of TCP parameters. However, as far as we know, *TCPTuner* is the first GUI tool of its kind to allow modification of congestion control parameters themselves on-the-fly.

## 3. TCPTUNER
The *TCPTuner* kernel module is designed to be loaded as the main kernel module for congestion control so that it serves as the algorithm that governs traffic on the machine. Because of its widespread use and its place as the default congestion control algorithm in Linux and OSX, we chose to clone and expose the parameters of CUBIC. Once the module is loaded, the *TCPTuner* GUI (Figure 1) allows a user to modify congestion control parameters during runtime. This section gives a high-level description of CUBIC and implementation details of *TCPTuner*. Currently, *TPCTuner* only runs on Linux.

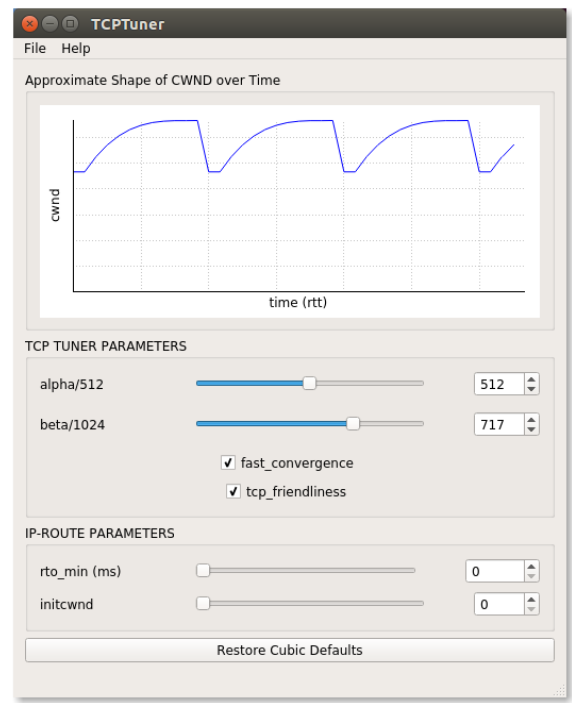

**Figure 1: TCPTuner GUI**

## 3.1 CUBIC Congestion Control

The internet today needs to support fast transfer of large volumes of data. However, traditional AIMD algorithms perform poorly due to their slow response times in fast, long distance networks (those with high bandwidth-delay products). The goal of CUBIC is to address this issue while maintaining fairness, stability, and compatibility with traditional TCP congestion control.

The most obvious difference between CUBIC and AIMD-based algorithms is CUBIC's window growth function. CUBIC does not have a strict constant increase on the receipt of ACKs, but instead finds a target window size and increases towards that target. The target window size, $W$, is a function of elapsed time and the maximum window size reached prior to the last packet loss event, $W_{max}$. Specifically, $W$ is defined by the function:

$$W_{cubic}(t) = C(t - K)^3 + W_{max} \quad (1)$$

Where $C$ is a scaling constant, $t$ is the elapsed time from the last window reduction, and where:

$$K = \sqrt[3]{\frac{W_{max}\beta}{C}} \quad (2)$$

With this growth function, flows with shorter RTTs (round-trip times) will not grow more quickly than those with long RTTs, unlike traditional AIMD algorithms, or even its predecessor, BIC. Instead of the traditional TCP sawtooth seen with AIMD algorithms, CUBIC's window growth function can be visualized as shown in Figure 2.

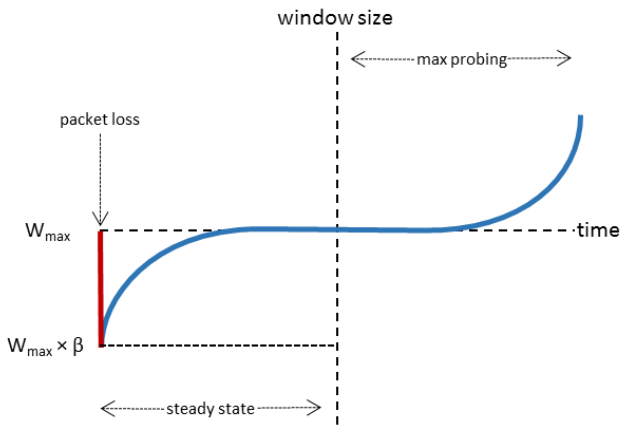

**Figure 2: TCP CUBIC Growth Behavior**

TCP CUBIC includes three parameters which affect window growth. First, *beta* (shown as β in Figure 2) controls how much the window is multiplicatively decreased. The default behavior in CUBIC is to have *beta* be 0.7 which corresponds to reducing the window by 30% on a loss event. Second, CUBIC includes a *fast convergence* parameter that modifies the value of $W_{max}$ when a loss event occurs. When *fast convergence* is disabled, $W_{max}$ is set to the value of the congestion window at the time of the loss event. If *fast convergence* is enabled, $W_{max}$ is scaled down to be approximately 85% of the congestion window when the loss event occurred. Third, *tcp friendliness* allows CUBIC to grow in the same way as standard TCP in short RTT environments. Both *fast convergence* and *tcp friendliness* are enabled by default in Linux.

In addition to exposing these three parameters, we add *alpha* (α) which scales the value of $W_{max}$ by setting:

$$W_{max} = W_{max} * \alpha \quad (3)$$

This allows *TCPTuner* to adjust how quickly the window grows by inducing a longer period of very fast growth after a reduction, or by causing the window to remain reduced until CUBIC enters the max probing state (see Figure 2).

One major challenge we found when working with TCP CUBIC were inconsistencies between descriptions and implementation of the CUBIC algorithm. There are two major papers that describe CUBIC: the original CUBIC draft introduced in 2004 [9] and an updated paper by the same authors in 2008 [3]. These two papers are unclear in their definition of $K$. Both the original and more recent paper show Equation 2, but the latter goes on to show:

$$K = \sqrt[3]{\frac{W_{max}\beta - cwnd}{C}} \quad (4)$$

in the description of the algorithm, which is how $K$ is actually implemented in the source code. Another subtle difference between theory and implementation is in the cubic growth function itself. The current Linux implementation of CUBIC adds an absolute value to the equation that is not present in either of the papers, defining window growth as:

$$W_{cubic}(t) = C|(t - K)|^3 + W_{max} \quad (5)$$

These nuances, along with other macros and optimizations, obfuscate the simplicity of the cubic function described in Equations 1 and 2. When creating the visual approximations discussed in Section 3.4, we referred directly to the CUBIC source code [6].

## 3.2 IP ROUTE Parameters

In addition to setting CUBIC's parameters, *TCPTuner* also allows users to modify two *ip route* parameters: *rto_min* and *initcwnd*. The first of these parameters is the minimum retransmission timeout or the minimum amount of time to wait for acknowledgment before triggering a timeout event. The second is the initial congestion window size. We chose to expose these parameters because, even though they are not part of the TCP module, they can directly affect the performance of TCP. For example, Dukkipati et. al. found that increasing TCP's *initcwnd* improved the average latency of HTTP responses [10]. Similarly, the loss recovery mechanism of TCP can be source of latency, with the default retransmission timeout set at 1 second [11]. In some cases, decreasing *rto_min* can improve responsiveness on unreliable networks.

### 3.3 Pluggable Congestion Module & GUI

Starting from kernel version 2.6.13, Stephen Hemminger incorporated a new architecture called *pluggable congestion modules*, which eliminated the need to recompile the kernel to experiment with new TCP algorithms [4][5]. This framework allows congestion control modules to be switched and loaded dynamically without recompilation. Each module provides an interface which serves as hooks that provide access to the module's code. All congestion control algorithms are required to define *cong_avoid* and *ssthresh*. Any other methods are optional. CUBIC is implemented as a pluggable congestion control module [3].

The *TCPTuner* pluggable congestion control module is a clone of CUBIC [6], with the addition of an *alpha* module parameter. All writable module parameters are accessible through the *sysfs* feature of Linux, which was put into place in the 2.6 kernel [7]. *Sysfs* is a virtual filesystem that provides a structured way to expose system information and attributes to user-space through virtual files. The subdirectory of /sys/ that is relevant for pluggable congestion modules is the module/ directory, which contains subdirectories for each module that is currently loaded into the kernel. The name of each module directory is the name of the module itself. If a module has parameters, they are stored in the parameters/ subdirectory. For example, the value of *TCPTuner*'s *fast convergence* parameter could be found by navigating to /sys/module/tcp_tuner/parameters/fast_convergence.

With the kernel module in place, users can manipulate various parameters through the *TCPTuner* GUI shown in Figure 1, which is written in C++ using the Qt5 framework [12]. The *sysfs* interface allows the *TCPTuner* GUI to read and write to each parameter's corresponding virtual file in order to set new values in the kernel module itself. In order to set *ip route* parameters, the GUI simply utilizes *ip route* commands for reading and setting parameters as one would normally do in a terminal.

Also note that the GUI shows values of *alpha* and *beta* being divided by 512 and 1024, respectively. This is because the CUBIC implementation divides *beta* by 1024. To maintain consistency and preserve the simplicity of integers, we have aligned the GUI to match. Thus, to have a multiplicative decrease of 50%, the *beta* slider must be set to 512. Currently the values of *alpha* are limited to the range (0, 2], while *beta* is limited to (0, 1].

### 3.4 Visualizing CWND

The *TCPTuner* GUI also includes a visualization of the approximate shape of the congestion window over time. The purpose of this visualization is to give the user a general sense of how manipulating *alpha* and *beta* will affect the cubic growth function. The visual makes the assumptions that there is only a single slow, *fast convergence* and *tcp friendliness* are both disabled, and that packet losses are only caused by overflowing a tail-drop queue. In the future, this visualization could be enhanced to give a more accurate estimate of actual congestion window sizes, but for our purposes, this approximation is sufficient.

Using the simulation environment described in Section 4.1.1, we measured the actual congestion window of a TCP sender for several values of *alpha* and *beta*, with *fast convergence* and *tcp friendliness* disabled. These measurements were sampled using the *getsockopts* system call at a rate of 5 samples per second. The results are shown in Figure 4. These experiments show that our simple model adequately serves as an approximation of the actual shape of a sender's congestion window based on our assumptions.

## 4. EXPERIMENTATION
### 4.1 Simulation
#### 4.1.1 Simulation Environment and Setup

In order to simulate and view the behavior of *TCPTuner*, we used the MahiMahi framework [8] on a machine running Ubuntu Linux 15.10. We simulated a 12Mbps uplink with an 80ms RTT and a 120000-byte tail-drop queue. This setup allowed us to monitor the throughput of multiple TCP flows running different protocols or the throughput and delay of a single flow. We only monitored the uplink from client to server. Clients sent a continuous stream of 1KB packets as quickly as the congestion control algorithm allowed.

#### 4.1.2 Inter-Protocol Performance

To start, we tested two TCP flows sharing the 12Mbps bottleneck link. One of the flows used standard TCP CUBIC. The other used the *TCPTuner* module, which allowed us to experiment with several values of *alpha* and *beta*. Figure 3 shows a CUBIC flow competing with a new *TCPTuner* flow with *alpha* and *beta* both set to 1. We see that with no multiplicative decrease, the *TCPTuner* flow quickly takes over the bottleneck link and starves the default CUBIC flow of bandwidth.

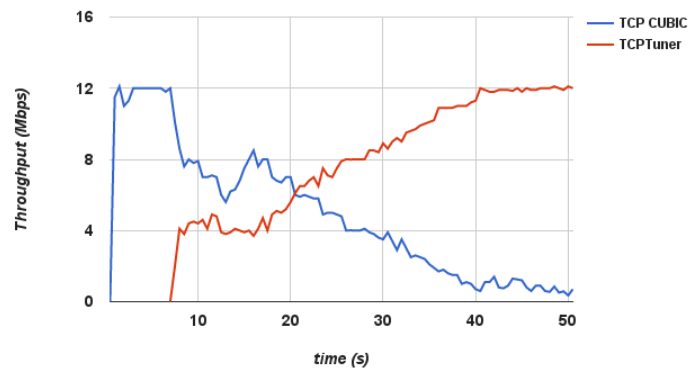

**Figure 3: CUBIC vs TCPTuner: alpha = beta = 1**

Figure 5 shows the same setup with a *beta* value of 0.25. In this case, we see that the CUBIC flow receives more bandwidth on average than the *TCPTuner* flow.

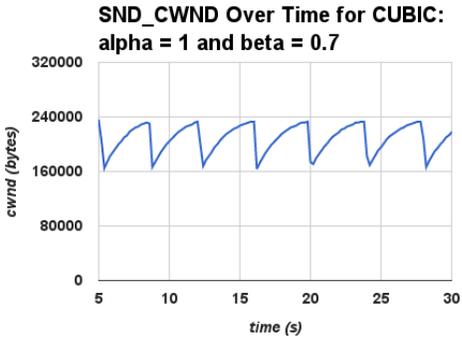
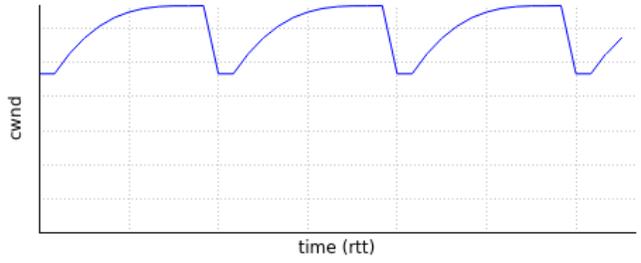
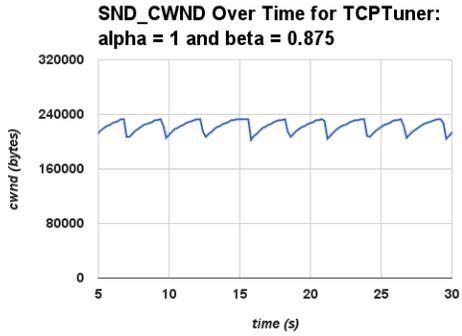
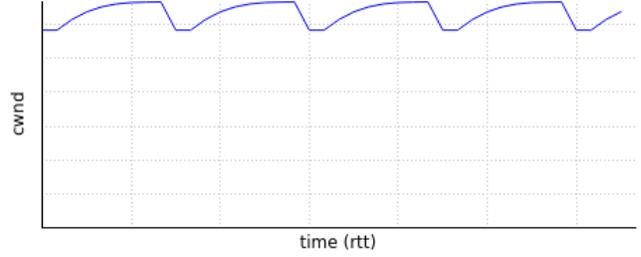
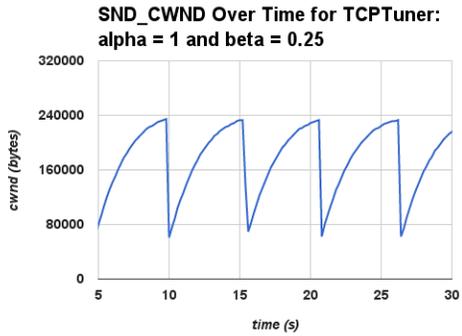
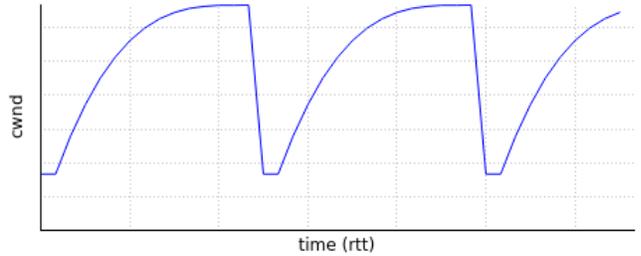
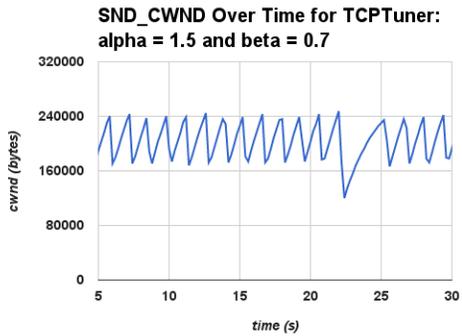
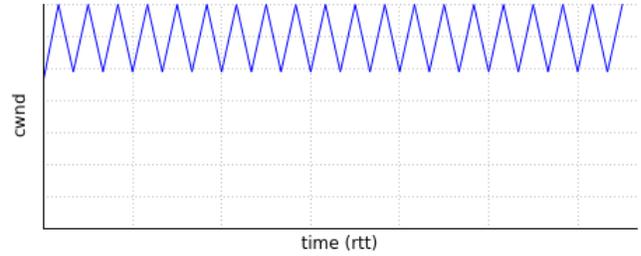
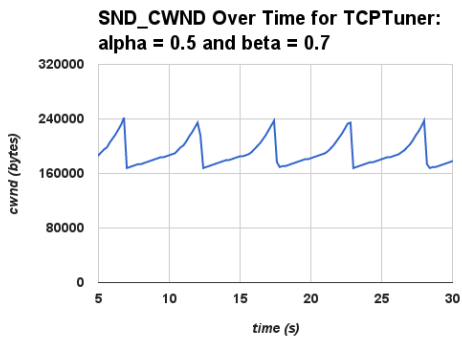
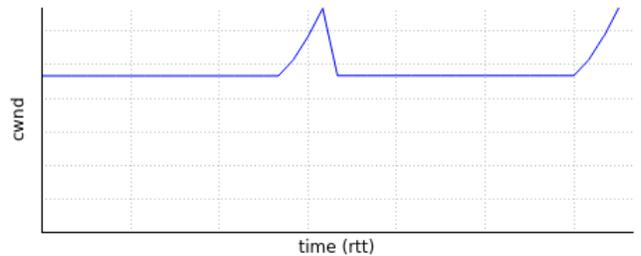

Figure 4: Comparison of Measured Congestion Window (left side) and the GUI Approximation (right side)

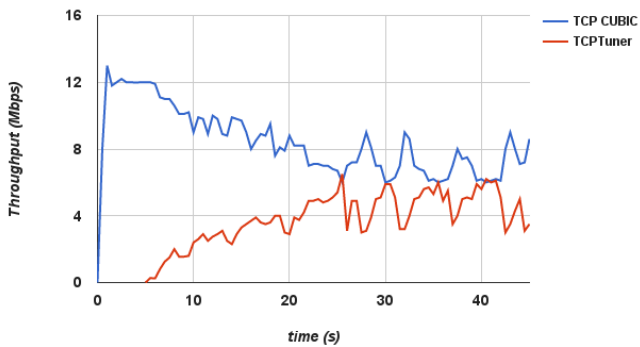

**Figure 5: CUBIC vs TCPTuner: alpha = 1, beta = 0.25**

Modifying *alpha* has similar, but less dramatic effects with *beta* held at the default value. With an *alpha* value of 2, we found that the *TCPTuner* flow used more bandwidth on average than the CUBIC flow, but did not monopolize the bottleneck link. Similarly, lowering the value of *alpha* resulted in the CUBIC flow receiving more bandwidth.

### 4.1.3 Intra-Protocol Performance

We then performed a similar experiment with two *TCPTuner* flows. Figure 6 shows result of one *TCPTuner* flow fully utilizing the bottleneck link and then introducing a second flow where both flows have their *alpha* and *beta* set to 1.

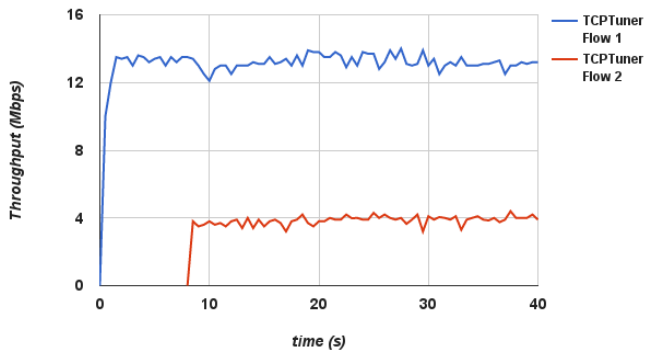

**Figure 6: Two TCPTuner Flows: alpha = beta = 1**

Notice that the two flows reach a steady state where the new flow uses approximately 4Mbps. However, the original flow still uses over 12Mbps, suggesting that the bottleneck link is heavily oversubscribed. Consequently, this experiment suggests that if many clients used *TCPTuner* with a *beta* value of 1, the productive throughput of the network would be reduced due to oversubscription of bottleneck links. We also see that a new *TCPTuner* flow will not receive a fair share of the bottleneck if a previous *TCPTuner* flow has already monopolized the link. If both flows have equal parameter values and *beta* is less than 1, fairness is achieved.

## 4.2 Real-world Performance

Next, we performed a file transfer experiment using *scp* on a 12Mbps simulated link with 1% probability of a packet loss and a 50ms RTT. We transferred a 1MB file using varying values of *beta*. As shown in Figure 7, increasing *beta* resulted in reduced transfer times. We also noticed that despite tuning *beta*, *scp* still underperformed the theoretical limit of transferring a 1MB (~8Megabit) file in under 1 second for a 12Mbps link.

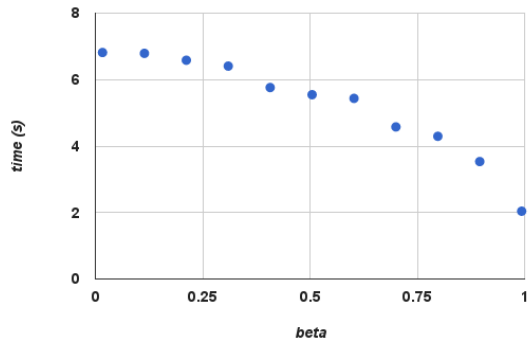

**Figure 7: Simulated SCP Transfer Time vs Beta**

We then performed another *scp* experiment (Figure 8) transferring a 50MB file from Stanford to an Amazon EC2 VM hosted in Sydney, Australia. The RTT to the server was approximately 265ms. Adjusting *beta* appears to improve performance, though not as significantly as our simulation.

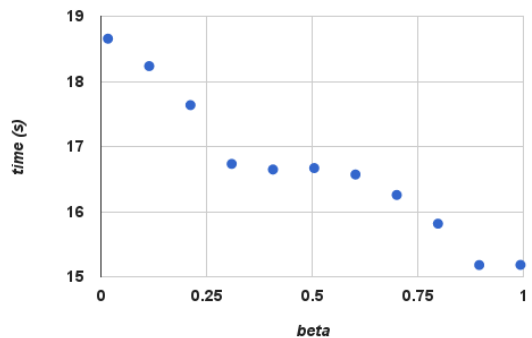

**Figure 8: SCP Transfer Time vs Beta (Stanford to Australia)**

When transferring the same 50MB file from Stanford to an Amazon EC2 VM hosted in Oregon, we found that *beta* had no influence on file transfer time. The RTT to this server was approximately 25ms. This suggests that the benefits that a single user can gain from adjusting *beta* depends on the characteristics of the network that they are using. In this case, it may be that there were not enough packet loss events for the modified *beta* to provide any significant advantages.

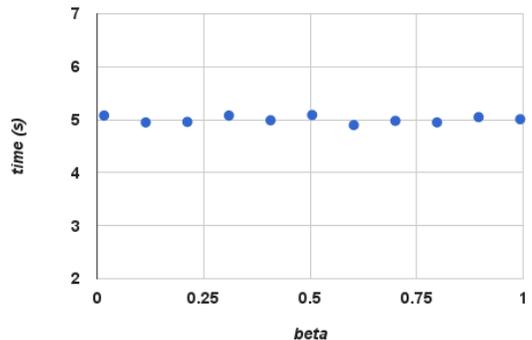

**Figure 9: SCP Transfer Time vs Beta (Stanford to Oregon)**

## 5. CONCLUSION

We present *TCPTuner*, a kernel module based on TCP CUBIC and a GUI that allows users to modify congestion control parameters dynamically during runtime. As far as we know, *TCPTuner* is the first software package of its kind. We describe its implementation and provide several experimental results that were enabled by *TCPTuner*. For example, we show that a TCP flow that does not multiplicatively decrease can quickly monopolize the throughput of a bottleneck link.

*TCPTuner* can provide value to researchers, industry, educators, and individuals. We hope that future congestion control related experiments can be done using *TCPTuner*. In addition, we envision *TCPTuner* being used as a teaching tool, allowing both instructors and students alike to see the effects of changing congestion control parameters in realtime.

## 6. ACKNOWLEDGMENTS

We would first and foremost like to thank Stanford Professor Keith Winstein for his support and advice throughout this project. We would also like to acknowledge the constructive feedback offered by the students of CS344G at Stanford University.